\documentclass[a4paper]{jpconf}
\usepackage{graphicx}
\begin{document}
\title{Azimuthal anisotropy ($v_{2}$) of high-p$_{T}$ $\pi^{0}$ and direct $\gamma$ in Au+Au collisions at
$\sqrt{s_{NN}}$ = 200 GeV }
\author{Ahmed Hamed (STAR Collaboration)}
\address{Texas A$\&$M University, College Station, USA}
\ead{ahamed@tamu.edu}

\begin{abstract}
Preliminary results from the STAR collaboration of the azimuthal anisotropy $(v_{2})$ 
of $\pi^{0}$ and direct photon ($\gamma_{dir}$) at high transverse momentum (p$_{T}$) from Au+Au collisions 
at center-of-mass energy $\sqrt{s_{_{NN}}}=200$~GeV are presented. 
A shower-shape analysis is used to select a sample free of direct photons ($\pi^0$) and a sample rich in direct photons
$\gamma_{rich}$. The relative contribution of background in the $\gamma_{rich}$ sample is determined 
assuming no associated charged particles nearby $\gamma_{dir}$. 
The $v_{2}$ of direct photons 
($v_{2}^{\gamma_{dir}}$) at mid-rapidity ($|\eta^{\gamma_{dir}}|<1$) and high p$_{T}$ ($8< p_{T}^{\gamma_{dir}}<16$~GeV/$c$) 
is extracted from those of $\pi^{0}$ and neutral particles measured in the same kinematic range. 
In mid-central Au+Au collisions (10-40$\%$), the $v_{2}$ of $\pi^0$ ($v_{2}^{\pi^{0}}(p_{T})$) and 
charged particles ($v_{2}^{ch}(p_{T})$) are found to be $\sim$ 0.12 and nearly independent of p$_{T}$. The measured
$v_{2}^{\gamma_{dir}}(p_{T})$ is 
positive finite and systematically smaller than that of $\pi^{0}$ and charged particles by a factor of $\sim$ 3. 
Although the large $v_{2}^{\pi^{0}}$ at such high p$_{T}$ might be partially due to the path-length dependence of energy loss,
the non-zero value of $v_{2}^{\gamma_{dir}}$ indicates a bias of the reaction plane determination due to the presence of jets in the
events. Systematic studies are currently in progress.
\end{abstract}
\section{Introduction}
The azimuthal distribution of the produced particles in heavy-ion collisions is expected
to be sensitive to the initial geometric overlap of the colliding nuclei, and would result in
anisotropic azimuthal distributions with respect to the reaction plane.
The standard method to quantify the azimuthal anisotropy is to expand the particle azimuthal 
distributions in a Fourier series $\frac{dN}{d\phi} = \frac{1}{2\pi} [ 1 + \sum_{n} 2v_{n}\cos(n(\phi_{p_{T}} - \phi_{RP}))]$,
where $\phi_{p_{T}}$ is the azimuthal angle of the produced particle with certain value of 
p$_{T}$, $\phi_{RP}$ is the azimuthal angle of the reaction plane, and $v_{n}$ is the coefficient of the
$n^{th}$ harmonic. The $2^{nd}$ Fourier moment ($n=2$) is referred to as the ``elliptic flow" parameter $v_{2}$ and its differential form 
is given by
\begin{equation}
v_{2}(p_{T}) = \langle e^{2i(\phi_{p_{T}}-\phi_{RP})} \rangle = \langle \cos 2(\phi_{p_{T}}-\phi_{RP})\rangle
\label{eq:TSP}
\end{equation}
where the brackets denote statistical averaging over different events.

While RHIC data show large amount of elliptic flow as predicted by the hydrodynamic models at low p$_{T}$, the results
at high p$_{T}$ deviate strongly from the hydrodynamic predictions as is expected~\cite{STAR_white}.  
The medium-induced radiative energy
loss of partons (jet-quenching) has been proposed as the source for the large observed azimuthal anisotropy at high p$_{T}$,
due to the path-length dependence of the
parton energy loss~\cite{Shuryak}. The STAR results~\cite{STAR1} show the amount of $v_{2}$ at high p$_{T}$ is 
larger than the predicted values 
by pure jet-quenching models. 
Although recent measurements by PHENIX~\cite{PHENIX0} show the produced $\pi^{0}$'s in-plane outnumber those 
produced out-of-plane which may be consistent with the path-length dependence of energy loss, the reaction plane determination
might have remaining bias toward the direction of the produced jets.
STAR results show insensitivity of the path-length-dependence of energy loss~\cite{STAR2} at high p$_{T}$ through
a comparison between $\gamma_{dir}$-charged particles and $\pi^{0}$-charged particles azimuthal correlations.

The $v_{2}$ measurement of direct photons would help to 
assess any remaining bias in the reaction plane determination leading to a positive $v_{2}$ signal.
Furthermore the $v_{2}^{\gamma_{dir}}$ would give additional information to help disentangle the various 
scenarios of direct photon production 
through the expected opposite contributions to the $v_{2}$~\cite{Theory1,Theory2,Theory3,Theory4}, and therefore
could help to confirm the observed binary scaling of the direct photon$~\cite{PHENIX2}$. 
\section{Analysis and Results}
\subsection{Electromagnetic neutral clusters}
The STAR detector is well suited for measuring azimuthal angular correlations 
due to the large coverage in pseudorapidity ($|\eta|<1$)
and full coverage in azimuth ($\phi$). 
While the Barrel Electromagnetic
Calorimeter (BEMC)~\cite{STAR_BEMC} measures the 
electromagnetic energy with high resolution, the Barrel Shower Maximum Detector (BSMD) provides fine spatial 
resolution and enhances the rejection power for the hadrons. The Time Projection Chamber (TPC)~\cite{STAR_TPC} identifies 
charged-particles, measures its momenta, and allows for a charged-particle veto cut with the BEMC matching.   
Using the BEMC to select events (\textit{i.e.} ``trigger") with high-$p_{T}$ $\gamma$,
the STAR experiment collected an integrated 
luminosity of 535~$\mu$b$^{-1}$ of Au+Au 
collisions in 2007. 
In this analysis, events having primary vertex within $\pm 55$ cm 
of the center of TPC along the beamline, and
with at least one electromagnetic cluster  
with $E_T > 8$~GeV are selected. More than 97$\%$
of these clusters have deposited energy greater than 0.5 GeV in each layer
of the BSMD. A trigger tower is rejected if it has a track 
with $p > 3.0 $~GeV/$c$ pointing to it, which reduces the number of the electromagnetic clusters by
only $\sim 7$\%. 
\subsection{$v_{2}$ of neutral and charged particles}
The $v_{2}$ is determined using the standard method (Eq. 1), which 
correlates each particle with the event plane determined from the full event minus
the particle of interest. The event plane is determined by 
\begin{equation}
\phi_{RP} = \frac{1}{2} \tan^{-1} (\frac{\sum_{i}\sin(2\phi_{i})}{\sum_{i}\cos(2\phi_{i})} )
\end{equation}
where $\phi_{i}$ are the azimuthal angles of all the particles used to define the event plane. In this analysis,
the charged-track quality criteria are similar to those 
used in previous STAR analyses~\cite{flow}. The event plane is measured 
twice: 1) using all the selected tracks inside the TPC (full-TPC), 
and 2) using the selected tracks in the opposite pseudorapidity side to the particle of interest (off-$\eta$) 
in order to reduce the ``non-flow" contributions (azimuthal correlations not related to the reaction plane). 
Since the event plane is only an approximation to the true reaction plane, 
the observed correlation is divided by the event plane resolution. The event plane resolution is estimated
using the sub-event method in which the full event is 
divided up randomly into two sub-events (for full-TPC and off-$\eta$ separately) as described in~\cite{flow2}. Biases due to the finite acceptance of the detector, which cause the 
particles to be azimuthally anisotropic in the laboratory system are removed 
according to the method in~\cite{shift}.

Figure 1 (left panel) shows the $v_{2}$ of  
charged particles ($v_{2}^{ch}$) at low p$_{T}$ using the event plane method (off-$\eta$)  
compared to previous STAR measurements~\cite{flow}, and 
the $v_{2}$ of the charged and neutral particles using the full-TPC and off-$\eta$ event plane methods at high p$_{T}$. 
At low p$_{T}$ the $v_{2}^{ch}$(off-$\eta$) is smaller than the $v_{2}$ using the full TPC and agrees well with 
the $v_{2}$$\{4\}$ (4-particle cumulant) method, in which 
the contribution of the non-flow is expected to be small. 
At high p$_{T}$ the two different
methods (full TPC and off-$\eta$) for the reaction plane measurements give similar results (both for the 
charged and neutral particles separately), which might 
indicate
the dominance of non-flow contributions in $v_{2}$ measurements. 
The $v_{2}(p_{T})$ of the
neutral particles show systematically smaller values than those of charged particles 
due to the $\gamma_{dir}$ contributions. 
\begin{figure}
\begin{center}
   \resizebox{140mm}{155pt}{\includegraphics{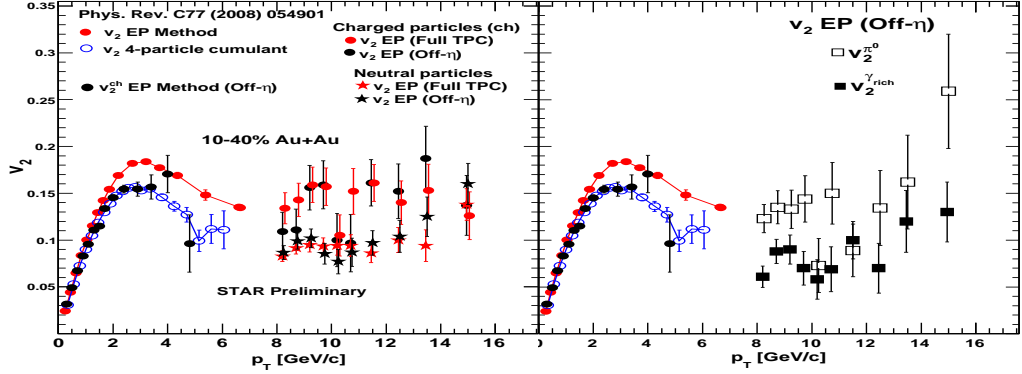}} 
        \caption{(Color online) For p$_{T} < 6$~GeV/$c$, both panels show previous STAR measurements~\cite{flow} of $v_{2}$
	as a function of p$_{T}$ for charged particles with $|\eta| <$ 1 in 10-40\% $Au+Au$
	collisions at $\sqrt{s_{_{NN}}}=200$~GeV using the Event-Plane method (closed red circles), and the 4-particle cumulant
	method (open circles). Also $v_{2}$ for charged particles ($|\eta| <$ 1) using off-$\eta$ event plane method is shown in closed black
	circles (this analysis). For p$_{T} > 6$~GeV/$c$: (left panel) $v_{2}$ of
	charged particles (red and black circles)
	and $v_{2}$ of neutral particles (red and black stars) using the full TPC and off-$\eta$ event plane methods
	respectively; (right panel) $v_{2}$ of $\pi^{0}$ and $\gamma_{rich}$ (open and closed squares, respectively)
	using the off-$\eta$ method. 
	}
    \label{fig:corrfunc}
    \end{center}
\end{figure} 
\begin{figure}
\begin{center}
   \resizebox{140mm}{155pt}{\includegraphics{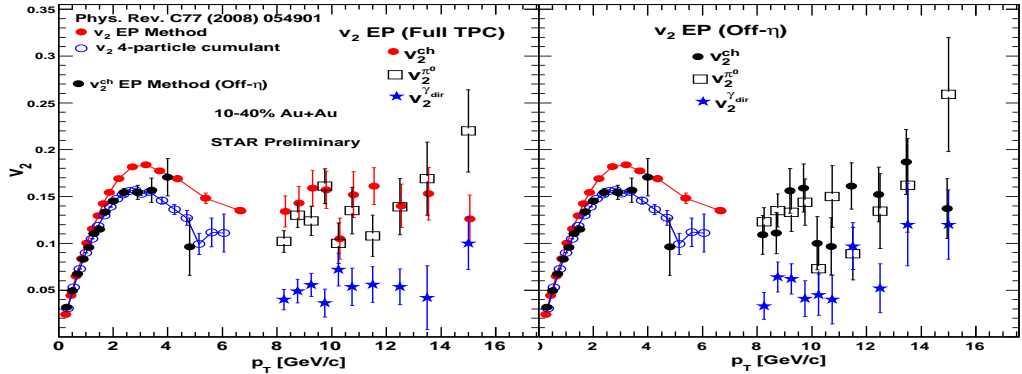}} 
        \caption{
	(Color online) For p$_{T} < 6$~GeV/$c$, both panels show measurements as in Fig. 1.
        For p$_{T} > 6$~GeV/$c$, 
	both panels show $v_{2}$ of
	charged particles, $\pi^{0}$, and $\gamma_{dir}$ (circles, squares, stars respectively) using the full TPC (left
	panel) and using off-$\eta$ methods (right panel). 
	}
    \label{fig:corrfunc}
     \end{center}
\end{figure} 
\subsection{Transverse shower profile analysis}
A crucial part of the analysis is to discriminate between showers from $\gamma_{dir}$ and 
two close $\gamma$'s from high-$p_{T}$ $\pi^{0}$ symmetric decays. 
At $p_T^{\pi^0} \sim 8$~GeV/$c$, the angular separation between the two $\gamma$'s 
resulting from a $\pi^{0}$ decay is small, but a $\pi^{0}$ shower is generally broader than a single $\gamma$ shower. 
The BSMD is capable of 
$(2\gamma$)/$(1\gamma)$ separation up to $p_T^{\pi^0} \sim 20$~GeV/$c$ due 
to its high granularity. 
The shower shape is quantified as the cluster energy, 
measured by the BEMC, normalized by the position-weighted energy moment, 
measured by the BSMD strips~\cite{STAR2}.
The shower profile cuts were tuned to obtain a nearly $\gamma_{dir}$-free 
($\pi^{0}_{rich}$) sample and a sample rich in $\gamma_{dir}$ ($\gamma_{rich}$). 
Since the shower-shape analysis 
is only effective for rejecting two close $\gamma$ showers, the $\gamma_{rich}$ sample 
contains a mixture of direct photons and contamination from 
fragmentation photons ($\gamma_{frag}$) and photons from asymmetric hadron ($\pi^0$ and $\eta$) decays.

Figure 1 (right panel) shows the $v_{2}$ of $\gamma_{rich}$ sample 
and $v_{2}$ of the sample free of direct photons ($\pi^{0}$) 
using the off-$\eta$ event plane method. As expected, the  $v_{2}$ of $\gamma_{rich}$ sample is smaller than that of
$\pi^{0}$, while the $v_{2}^{\pi^{0}}$ is similar to $v_{2}^{ch}$ which indicates 
the $\pi^{0}$ sample identified by the
transverse shower-shape analysis, to be free of $\gamma_{dir}$. 
\subsection{v2 of direct photons}
The $\it v_{2}^{\gamma_{dir}}$ is given by:
\begin{equation}
v_{2}^{\gamma_{dir}}=\frac {v_{2}^{\gamma_{rich}}- {\cal{R}}v_{2}^{\pi^{0}}} {1-\cal{R}} 
\end{equation}
where $\cal{R}$=$\frac{N^{\pi^{0}}}{N^{\gamma_{rich}}}$, and the numbers of
$\pi^{0}$ and $\gamma_{rich}$ triggers
are represented by $N^{\pi^{0}}$ and 
$N^{\gamma_{rich}}$ respectively. 
The value of $\cal{R}$ is measured in~\cite{STAR2} and 
found to be 
$\sim 30\%$ in central Au+Au. 
In Eq. 3 all background sources for $\gamma_{dir}$ are assumed to have the same $v_{2}$ as $\pi^{0}$. 
Thus, excepting those $\gamma_{frag}$ that have no near-side yield, all other sources of $\gamma_{dir}$'s
background are removed. 

Figure 2 (left and right panels) shows the $v_{2}^{\gamma_{dir}}$ and $v_{2}^{\pi^{0}}$ compared to $v_{2}^{ch}$ at high p$_{T}$ using
the two different event-plane methods. 
While the $v_{2}^{\pi^{0}}$ and $v_{2}^{ch}$ are similar ($\sim$ 12$\%$), the $v_{2}^{\gamma_{dir}}$ 
is systematically lower than that of hadrons. 
The similarity of the $v_{2}$ results using the full-TPC and off-$\eta$ at high p$_{T}$, along with 
the non-zero value of $v_{2}^{\gamma_{dir}}$, indicate the non-flow contributions to the measured 
$v_{2}$.

\section{Conclusions}
The STAR experiment has reported the first $v_{2}^{\gamma_{dir}}$ at high-p$_{T}$ at RHIC. 
The non-zero value of $v_{2}^{\gamma_{dir}}$ is probably due to contributions from non-flow to the
standard method of $v_{2}$ measurements, where the pseudorapidity gap between the particle of interest and
the particles used for the reaction plane determination is smaller than two units in pseudorapidity. The positive value of 
$v_{2}^{\gamma_{dir}}$ demonstrates the negligible contribution of jet-medium photons~\cite{Theory2} and possible contributions 
of ${\gamma_{frag}}$~\cite{Theory1} to the direct photon productions over the covered kinematics range.

\end{document}